\documentstyle{article}

         \def\ba{\begin{array}}
         \def\ea{\end{array}}
         \def\be{\begin{equation}}
         \def\bea{\begin{eqnarray}}
         \def\eea{\end{eqnarray}}
         \def\ee{\end{equation}}

         \def\ds{\displaystyle}
\begin{document}
\hfill{ }

\hfill{hep-th/9610015  } 

\vskip 10 mm
\leftline{ \Large \bf
           Toda Theories as  Contractions of Affine Toda Theories        
                        } 
\vskip 10 mm
\leftline{ \bf
          A. Aghamohammadi ${}^{1,2,*}$,
          M. Khorrami ${}^{1,3,4}$, and
          A. Shariati ${}^{1,4}$
          }
\vskip 10 mm
{\it
\leftline{ $^1$ Institute for Studies in Theoretical Physics and Mathematics,
           P.O.Box  5531, Tehran 19395, Iran. }
\leftline{ $^2$ Department of Physics, Alzahra University,
             Tehran 19834, Iran }
\leftline{ $^3$ Department of Physics, Tehran University,
             North-Kargar Ave. Tehran, Iran. }
\leftline{ $^4$ Institute for Advanced Studies in Basic Sciences,
             P.O.Box 159, Gava Zang, Zanjan 45195, Iran. }
\leftline{ $^*$ E-Mail: mohamadi@irearn.bitnet}
  }
\begin{abstract}
Using a contraction procedure, we obtain Toda theories and their structures, 
from affine Toda theories and their corresponding structures. By 
structures, we mean the equation of motion, the classical Lax pair, the 
boundary term for half line theories, and the quantum transfer matrix. 
The Lax pair and the transfer matrix so obtained, depend nontrivially on 
the spectral parameter.
\end{abstract}

\vskip 10 mm
\noindent A Toda field theory is an integrable theory of $r$ scalar fields,
based on a semisimple Lie algebra \cite{OT}. The equation of motion of such
a theory is 
\be \label{m1}
-\Box  \rho_a +\sum _b C_{ab} e^{\rho_b}=0,    \ee
where 
\be C_{ab}:={2<\alpha_a,\alpha_b>\over <\alpha_b,\alpha_b>} 
           =:{2K_{ab} \over K_{bb}} \ee
is the Cartan matrix of the Lie algebra, $K_{ab}$ is its Killing form,
and $\alpha_a$'s are the simple roots of the Lie algebra. Now the necessary
and sufficient conditions for a matrix to be the Cartan matrix of a 
semisimple Lie algebra are
\\ 1) \hskip 1cm $ \ds{ C_{ab}={2<\alpha_a,\alpha_b>\over 
                   <\alpha_b,\alpha_b>} }, $ \\
where $ < \cdot , \cdot > $ denotes a nondegenerate inner product,
\\ 2) \hskip 1cm $ \ds{ C_{ab} \le 0,} \hskip 5mm a\ne b, $
\\ 3) \hskip 1cm $ \ds{ C_{ab} }$'s are integers,
\\ 4) \hskip 1cm the set of $ \alpha_a $'s are linearly independent.

It has been shown, however, that not all of these conditions are necessary
for the theory to be integrable; in fact, one can omit the last condition
without destroying the integrability of the theory \cite{OT,AFZ,MOP}. If 
$ \alpha_a $'s are linearly dependent, the corresponding theory is called
an affine Toda theory. 

The matrices satisfying the first three conditions have been classified
\cite{BMW}. It is shown that if no proper (and nonempty) subset of $\alpha_a$'s
exists which is orthogonal to others, then there is just one linear 
combination of $\alpha_a$'s which is equal to zero:
\be \label{m2} \sum_an^a \alpha_a = 0 \ee
and all of the coefficients of this linear combination can be chosen 
positive. So, eliminating any $\alpha_a$ from the system, reduces it
to the simple root system of an ordinary semisimple Lie algebra.
Therefore, to obtain an affine Cartan matrix one must add a vector to the
simple root system of an ordinary semisimple Lie algebra, in a way which
preserves the properties 1--3. One way of doing this is to add the lowest
root (the negative of the highest root) to the system. The affine Cartan
matrix so formed, is called untwisted and the coefficients $n^a$ are then
the Kac lables. For some Lie algebras there exist other ways of extension.
The affine Cartan matrices so formed are called twisted. There exists a 
complete classification of all of these extensions in the literature.

There are certain symmetry relations in some (affine) Dynkin diagrams.
People have used these relations to construct a(n affine) Lie algebra   
from another and the corresponding (affine) Toda theory from another one,
by the procedure called folding \cite{OT} (or, its generalisation, reduction
\cite{PKS}).

Faddeev has introduced another kind of transformation \cite{FAD}. He starts
from the sine-Gordon (untwisted affine $A_1$ Toda) theory, shifts the 
sine-Gordon field to infinity, and obtains (by a suitable scaling of the 
parameters of the theory) the Liouville ($A_1$ Toda) theory.  In this way,
he obtains a new quantum Lax operator for the Liouville theory, which
depends nontrivially on the spectral parameter.

We extend this transformation (which we call it contraction) 
to all of (untwisted or twisted) affine
Toda theories, and obtain  quantum Lax operators, which depend nontrivially
on the spectral parameters, for Toda theories. It is important to mention
that there already exists a Lax operator for any Toda theory. These Lax
operators depend, however, trivially on the spectral parameter; that is,
the determinant of these Lax operators do not depend on the spectral 
parameter \cite{FAD}. 

We also show that this contraction works for the case
of Toda theories on the half line as well; that is, one can construct a
Toda theory on the half line from the corresponding affine Toda theory,
which is already well known.

Starting from (\ref{m1}) and using (\ref{m2}) it is easily seen that
\be \Box \sum _an^a \rho_a=0. \ee
This means that, there is a linear combination of the fields 
$\sum_an^a \rho_a$ which is free.
One can expand the field $\rho_a$ around the equilibrium position 
$\hat \rho_a$: 
\be \sum_b C_{ab}e^{\hat \rho_b}=0 \ee
or 
\be \sum_b K_{ab}{2e^{\hat \rho_b}\over <\alpha_b, \alpha_b>}=0. \ee
Comparing the above equation with (\ref{m2}), we get
\be {2e^{\rho_b}\over <\alpha_b, \alpha_b>}=m^2 n^b, \ee
which leads to
\be -\Box  (\rho_a-\hat  \rho_a) +m^2 \sum _b n^bK_{ab} 
e^{\rho_b-\hat \rho_b}=0. \ee
Defining $\phi_a:= \rho_a- \hat\rho_a$, the equation of motion of $\phi_a$
is
\be -\Box  \phi_a +m^2 \sum _b n^bK_{ab} e^{\phi_b}=0. \ee
Note that according to (\ref{m2}), the combination $n^a \phi_a$ is free.
One can consistently set this combination equal 
to zero. $\phi_0$ is then not an independent field.
\be \label{m3} n^0\phi_0=- \sum_{\mu =1}^r n^{\mu}\phi_{\mu}. \ee
Throughout this article, indices $\{ a,b,\cdots\}$ run from 0 to $r$ and 
$\{ \mu,\nu,\cdots\}$ from 1 to $r$.
The equation of motion of $\phi:=\alpha^{\mu}\phi_{\mu}$ is
\be -\Box  \phi +m^2 \sum _b n^b\alpha_b e^{\phi_b}=0, \ee
and the equation of motion of $\phi^{\mu}$ is
\be \label{m4} -\Box  \phi^{\mu} +m^2  n^{\mu}( e^{\phi_{\mu}}-e^{\phi_0})=0,
\ee
where raising and lowering of the indeces are performed by the metric
$K_{\mu \nu}$ which is invertible.

Consider the equation (\ref{m1}). Shifting the $\rho_a$ fields 
\be \ba{c} \rho_a\to \rho_a+\xi_a  \cr
     \xi_I\to \infty , \quad 0 \le I \le s-1 \quad \quad \quad 
     \xi_i=0, s \le i \le r \ea \ee
we say that $\rho_I$'s are contracted. We use capital indices for contracted 
fields, the fields shifted to infinity, and small indices for the remainings. 
Also note that the ordering of the indeces is unimportant: one can
eliminate any of the fields $\phi_a$ in (\ref{m3})--(\ref{m4}). We choose
the eliminated field to be a contracted one.
It is obvious that the contracted fields disappear from the second term 
in the equation (\ref{m1}):
\be \label{m5} -\Box  \rho_a +\sum _i C_{ai} e^{\rho_i}=0. \ee
The above equation for the $\rho_j$'s is a Toda equation:
\be -\Box  \rho_j +\sum _i C_{ji} e^{\rho_i}=0,  \ee
where $C_{ji}$ is the Cartan matrix $C_{ab}$ in which the rows and 
coloumns corresponding to the contracted indices have been omitted. 
Eliminating some of $\alpha_a$'s makes the remainings 
linearly independent, and the remaining Cartan matrix is associated to a 
semisimple lie algebra. 
In Dynkin diagram, this contraction means that, the roots with contracted 
indices have been removed from the corresponding affine Dynkin diagram. 
In the original theory there were $r+1$ fields, a linear combination of them 
was free. If one contracts $s$ fields, $ 0\leq I \leq s-1 $, the fields
$\rho_i, \ \ \ s\leq i\leq r$ are Toda fields. We will show 
that  there are $s$ linear combinations of the fields which are free.
To show this, one should find vectors $p^a$ for which 
\be \label{m6} \sum_ap^aC_{ai}=0\ \ {\rm or}\ \  \sum_a p^aK_{ai}=0. \ee
Multiplying the equation (\ref{m5}) by $p^a$, one obtains
\be -\Box \sum_ap^a \rho_a =0. \ee
One of the solutions of (\ref{m6}) is 
\be p^{(0) a}=n^a. \ee 
The remaining $s-1$ solutions are 
\be p^{(I)a}=K^{Ia}, \ee
where  
\be \sum_{\nu}K^{\mu \nu}K_{\nu \rho}=
\delta^{\mu}_{\rho},\ \ {\rm and }\ \ K^{a0}=K^{0a}=0.  \ee
So the fields 
\be \sum_an^a\rho_a ,\ \ \rho^I :=\sum_{\mu}K^{I\mu}\rho_{\mu}, 
\ \ \ ( 1\leq I\leq s-1 )\ee
are free.
We want to use this method for obtaining boundary terms and Lax operators,
classical and quantum, of the Toda theories from the known ones for  
the affine Toda theories \cite{CDRS,MAC}, which are expressed in terms of 
$\phi$. 
So we study this contraction procedure in the language of $\phi$ fields.
For convinience, we use the primed indices for the fields and mass parameter 
before contraction:
\be -\Box {\phi '}_{\mu} + {m'}^2 \sum _b n^bK_{\mu b} 
e^{{ \phi '}_b}=0, \ee
\be -\Box  {\phi '}^{\mu} +{ m'}^2  n^{\mu}( e^{{\phi '}_{\mu}}-
e^{{\phi '}_0})=0. \ee
We shift the fields ${\phi '}_b= \phi_b +\zeta_b$, but in this case 
$\zeta$'s 
are not indepndent: 
\be n^b\zeta_b=0 \ee
or
\be n^i\zeta_i+n^I\zeta_I=0. \ee
We choose $\zeta_i=\zeta$ and $\zeta_I= \zeta '$ for all $i$'s and $I$'s.
Sending $ \zeta ' \to -\infty$ , $\zeta \to \infty$ and $ m'\to 0$ 
with the condition that $ {m'}^2e^{\zeta}=m^2$ remains finite,
one arrives at 
\be -\Box  \phi_{\mu} + m^2 \sum _i n^iK_{\mu i} e^{ \phi_i}=0. \ee
For the noncontracted indices $j$
\be -\Box  \phi_j + m^2 \sum _i n^iK_{j i} e^{ \phi_i}=0, \ee
and for the contracted indices $I$
\be -\Box  \phi^I=0. \ee
So the fields $\phi^I$ are free.

Now, consider the affine Toda theories on a half line. These theories have 
a boundary term $\cal B$, which is a function of the fields but not their
derivatives and represents the boundary conditions.
In \cite{CDRS} the generic form of the integrable boundary interaction 
is given by 
\be {\cal B}'= m'\sum_0^r A_a e^{\phi'_a/2}\ee
where the coefficients $A_a$, $a=0,\cdots r$ are a set of real numbers.
This term is a generalization of the results of the sine-Gordon theory.
Applying the contraction to the boundary term ${\cal B}'$, we obtain 
the boundary term $\cal B$ for the Toda theories which can be obtained 
from the affine ones via contraction.
\be {\cal B}= m\sum_i A_i e^{\alpha_i.\phi /2}.\ee

The same procedure can be repeated for the Lax operators. Classically, the 
field equation of affine Toda theories may be written as the zero curvature 
condition
\be[\partial_x+L_x, \partial_t+L_t]=0,\ee
where the Lax pair is \cite{MAC,OT,BOG}
\be\label{m9} L_x={1\over 2}\sum_{\mu} H_{\mu} \partial_t  {\phi '}^{\mu} + 
{m'\over 2}[\sum_{\mu} e^{{\phi'}_{\mu}/2}(E^+_{\mu}+E^-_{\mu})
+ e^{{\phi'}_{0}/2}(\lambda ' E^+_{0}+
{\lambda '}^{-1} E^-_{0})],\ee
\be\label{m10} L_t={1\over 2}\sum_{\mu} H_{\mu} \partial_x { \phi '}^{\mu} + 
{m'\over 2}[\sum_{\mu} e^{ {\phi'}_{\mu}/2}(E^+_{\mu}-E^-_{\mu})
+ e^{{\phi'}_{0}/2}(\lambda ' E^+_{0}-
{ \lambda '}^{-1} E^-_{0})],\ee
where $H_\mu$'s are the cartan generators, $E^{\pm}_\mu$'s the step 
operators corresponding to the simple roots, and $E^{\pm}_0$s are
the step operators associated with the added root. Contracting the fields 
and the parameters as below 
\be {\phi '}_0=\phi_0+{ \zeta '}, \quad 
 \phi_{\mu}=\phi_{\mu}+ \zeta\ee
\be m' e^{\zeta /2}=m \quad 
  m' e^{{ \zeta '} /2}{\lambda '}=m \lambda ,\ee
one arrives at the following relations for the Lax pair of the Toda 
equation:
\be L_x={1\over 2}\sum_{\mu} H_{\mu} \partial_t  \phi^{\mu} +{m\over 2}
[\sum_{\mu} e^{ \phi_{\mu}/2}(E^+_{\mu}+E^-_{\mu})
+ e^{ \phi_{0}/2} \lambda E^+_{0}],\ee
\be L_t={1\over 2}\sum_{\mu} H_{\mu} \partial_x  \phi^{\mu} +{m\over 2}
[\sum_{\mu} e^{\phi_{\mu}/2}(E^+_{\mu}-E^-_{\mu})
+ e^{ \phi_{0}/2}\lambda E^+_{0}].\ee
Note that this Lax pair of the Toda equation depends nontrivially on the 
spectral parameters. The spectral parameter dependence of the Lax pair 
(\ref{m9}) and (\ref{m10}) is nontrivial: there is no rescaling of the roots 
by which one can eliminate $\lambda$ \cite{MAC}. It can be easily shown 
that if one removes the parts proportional to $E_0^\pm$ from the Lax pair 
(\ref{m9}) and (\ref{m10}), there remains a Lax pair which gives the Toda 
field equation. This Lax pair, however, does not depend on the spectral 
parameter or, if one transports the spectral parameter to one of $E_\mu$'s, 
depends on the spectral parameter trivially. Faddeev has obtained a similar 
Lax pair, which depends nontrivially on the spectral parameter, for the 
Liouville theory, by contracting the Lax pair of the sine-Gordon theory.
Also note that this contraction procedure works for the quantum theory 
as well: the transfer matrix of the affine Toda theories, which has been 
obtained in \cite{MAC}, can be used quite similarly to obtain a tranfer 
matrix of the Toda theories, which depends nontrivially on the spectral 
parameter, that is, its determinant does depend on the spectral parameter.

\newpage


\begin{thebibliography}{99}
\bibitem{OT} Olive D., N. Turok; Nucl. Phys. {\bf B215} (1983) 470.
\bibitem{AFZ} Arinshtein A. E., V. A. Fateev, \& A. B. Zomolodchikov; 
              Phys. Lett. {\bf B87} (1979) 389.
\bibitem{MOP} Mikhailov A. V., M. A. Olshanetsky, \& A. M. Perelomov; 
              Commun. Math. Phys. {\bf 79} (1981) 473.
\bibitem{BMW} Berman S., R. Moody, M. Wonenburger; Indiana Univ. Math. J. 
              {\bf 21} (1972) 1091.
\bibitem{PKS} Patrik Khastgir S., R. Sasaki;        
              YITP-95-19, hep-th/9512158
\bibitem{FAD} Faddeev L. D., Tirkkonen O.;
              Nucl. Phys. {\bf B453} (1995) 647.
\bibitem{CDRS} Corrigan E., P. E. Dorey, R. H. Rietdjik and R. Sasaki;
               Phys. Lett. {\bf B333} (1994) 83.
\bibitem{MAC} Mackay N. J.;
              Mod. Phys. Lett. {\bf A9} (1994) 2353.
\bibitem{BOG} Bogoyavlensky, O. I.; Comm. Math. Phys. {\bf 51} (1976) 201.
\end{thebibliography}
\end{document}